\documentstyle[aps,preprint,epsfig]{revtex}
\tighten
\begin{document}
%
%
\title{The Electron Affinity of Li: A State Selective
Measurement}
\author{Gunnar Haeffler, Dag Hanstorp, Igor Kiyan\cite{byline}, 
Andreas E.~Klinkm\"uller, and Ulric
Ljungblad}
\address{G\"oteborg University and Chalmers University of Technology\\
Dept.~of Physics, S-412\,96 G\"oteborg, Sweden}
\author{David~J.~Pegg}
\address{Department of Physics, University of Tennessee\\
Knoxville Tennessee 37996, USA}
\date{\today}
%
%
\maketitle
\begin{abstract}\widetext
We have investigated the threshold of photodetachment of Li$^{-}$ leading to the
formation of the residual Li atom in the $2p\,^{2}\!P$ state. The excited
residual atom was selectively photoionized via an intermediate Rydberg state
and the resulting Li$^{+}$ ion was detected. A collinear laser-ion beam 
geometry enabled both
high resolution and sensitivity to be attained. We have demonstrated the 
potential 
of this state selective photodetachment spectroscopic method by improving the 
accuracy of Li
electron affinity measurements an order of magnitude. From a fit to the
Wigner law in the threshold region,
we obtained a Li electron affinity of $0.618\,049(20)$~eV.
\end{abstract}
\bgroup\draft\egroup
%
\narrowtext
\bibliographystyle{prsty}

\section{Introduction}
\label{intro}

The four-electron Li$^{-}$ ion
is interesting because of the
significant role played by electron correlation in the binding of the
outermost electron in this weakly bound system. The major contribution
to the correlation energy arises from the interaction of the two
valence electrons. Beyond the three-body H$^{-}$ system, the Li$^{-}$ ion is
most tractable to theory. In the frozen core approximation, for
example, the Li$^{-}$ ion becomes an effective three-body system consisting
of a pair of highly correlated electrons interacting weakly with an
inert core. Such a model lends itself well to semi-empirical model
potential calculations in which the potential experienced by the
valence electrons is obtained via spectral information on the
eigenvalue spectrum of the Li atom. Most calculations of the electron
affinity of Li to date are of this type, e.g.~Moccia et
al.~\cite{Moc_90} or Graham et al.~\cite{Gra_89} (and references therein).
Recently, however, accurate
measurements of this quantity have stimulated ab initio
calculations \cite{Chu_92_2,Fis_93} of comparable accuracy.

The most direct, and potentially the most accurate, method of
measuring electron affinities is to use the laser threshold
photodetachment (LTP) method \cite{Neu_85}. Here one records, as a function of
the wavelength of a tunable laser, the onset of production of either
photoelectrons or residual atoms in the vicinity of a detachment
threshold. 

To date, three LTP measurements of the electron affinity of Li have
been reported. 
The earliest such experiment was a crossed laser and ion beam experiment
by Feldman \cite{Fel_76}. He utilized an infrared laser to study the
Li$(2s)+\epsilon p$ photodetachment threshold. An accurate threshold
energy for a $p$-wave detachment was difficult to determine because the
cross section rises, according to the Wigner law \cite{Wig_48}, only slowly from zero.
Bae and Peterson \cite{Bae_85} used collinear laser and ion beams 
to investigate the
total photodetachment cross section around the Li($2p$) cusp situated at
the opening of the Li$(2p)+\epsilon s$ channel. From a careful
analysis of this sharp structure they obtained, as one result, an
electron affinity value of about the same accuracy as Feldman.
The measurement of Dellwo et al.~\cite{Del_92_2} was a
direct investigation of the resolved Li$(2p)+\epsilon s$ 
channel threshold using
photoelectron spectroscopy. In this experiment, however, Doppler 
broadening associated
with the use of crossed laser and ion beams limited the attainable 
energy resolution.

The electron affinity of Li determined in the present measurement
is an order of magnitude more accurate than previous LPT
measurements. 
We utilized resonance ionization \cite{Hur_79,Hur_88} combined with a 
collinear laser-ion
beam geometry to measure the threshold of the Li$(2p)+\epsilon s$
partial photodetachment cross section. The state selectivity of the
resonance ionization method leads to an excellent
signal-to-background ratio. This in turn enabled us to attain a resolution
limited only by the laser bandwidth of about 0.2~cm$^{-1}$. 
The
present threshold energy measurement clearly demonstrates the
potential of the method. The concept of combining collinear laser-ion
beam spectroscopy with resonance ionization detection was first
proposed by Kudriatsev and Letokhov \cite{Kud_82}
and later applied to isotope
detection measurements by the same authors \cite{Kud_88}.
Balling et al.~\cite{Bal_93} 
and Petrunin et al.~\cite{Pet_95}
have recently used the same technique in
photodetachment measurements.

\section{Experiment}
\label{experiment}

\subsection{Procedure}
\label{procedure}

The two-color 
state selective photodetachment experiment described in the present
paper is simple in concept. One laser of frequency
$\omega_{1}$ is used to photodetach Li$^{-}$ ions producing an 
excited Li atom and a free electron (Fig.~1). A second laser of
frequency 
$\omega_{2}$ resonantely photoexcites Li atoms left in the $2p$ state
to a Rydberg state which subsequently is field ionized. Hence, the 
entire process can be represented by the following steps: 

\begin{figure}
\epsfig{file=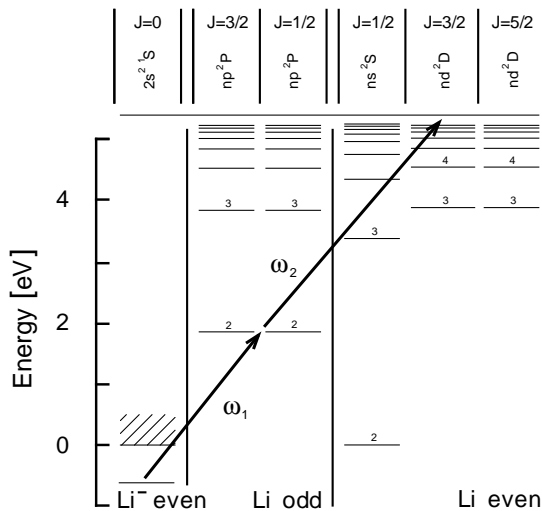, width=0.45\textwidth}
\caption{Excitation scheme: Selected bound states of
Li/Li$^{-}$ grouped according
to their
parity and total angular momentum.
The arrows indicate transitions induced in this experiment.}
\label{term}
\end{figure}

\begin{eqnarray}\label{stepone}
\mbox{Li}^{-}(2s^{2}\,^{1}\!S_{0}) + \hbar\omega_{1} &\rightarrow &
\mbox{Li}(2p\,^{2}\!P_{1/2,3/2}) +
\mbox{e}^{-}\quad ,\nonumber\\
\mbox{Li}(2p\,^{2}\!P_{1/2,3/2}) + \hbar\omega_{2} &\rightarrow &
\mbox{Li}(ns,nd)\quad , \nonumber\\
\mbox{Li}(ns,nd)&\leadsto &
\mbox{Li}^{+}(1s^{2}\,^{1}\!S_{0}) + \mbox{e}^{-}\quad ,
\end{eqnarray}
where $\leadsto $ denoted field ionization and 
Li($ns,nd$) corresponds to a highly excited Rydberg atom in either
a  $ns$ or $nd$ state.
State selectivity is accomplished in the resonant
ionization step since only Li($2p$) atoms can be ionized via the
intermediate Rydberg state. 
In this manner we 
were able to isolate a particular photodetachment channel, 
in this case
the Li($2p$) channel, and investigate the partial photodetachment cross 
section by measuring the yield of Li$^{+}$ ions. 

\subsection{Setup}
\label{setup}

The $^{7}$Li$^{-}$ ion beam was produced by charge exchange 
in a cesium  vapor cell
of a mass selected
Li$^{+}$ beam from a plasma ion source.
An ion current of typically a few nA was obtained in the interaction 
region. The beam energy was approximately 4~keV. 

In the interaction chamber (Fig. 2)  the negative ions interacted with
laser light in a region defined by two apertures with a diameter of 3~mm
placed 0.5~m apart.  The ions were deflected in and out of the laser
beam
by means of two electrostatic quadrupole deflectors whose symmetry axes
were perpendicular to the laser and ion beams. The ion current in the
interaction region was monitored with a Faraday cup placed after the
second
quadrupole deflector.

Rydberg atoms formed in the interaction region travel to the second
quadrupole where they are ionized by the same electric field 
that deflects the negative ion beam into the Faraday cup. 
Positive ions formed in this
process were deflected in the opposite direction into a positive ion
detector.
 In this detector the fast positive
ions
 impinged on a conducting glass  plate producing  secondary electrons that were
detected with a channel electron multiplier (CEM). A metal grid 
connected to
a voltage supply was placed between the glass plate and the CEM. This
made
it possible to  either allow or prevent the secondary electrons from
reaching the CEM. The detection efficiency of the positive ion detector
was
close to unity.

\begin{figure}
\epsfig{file=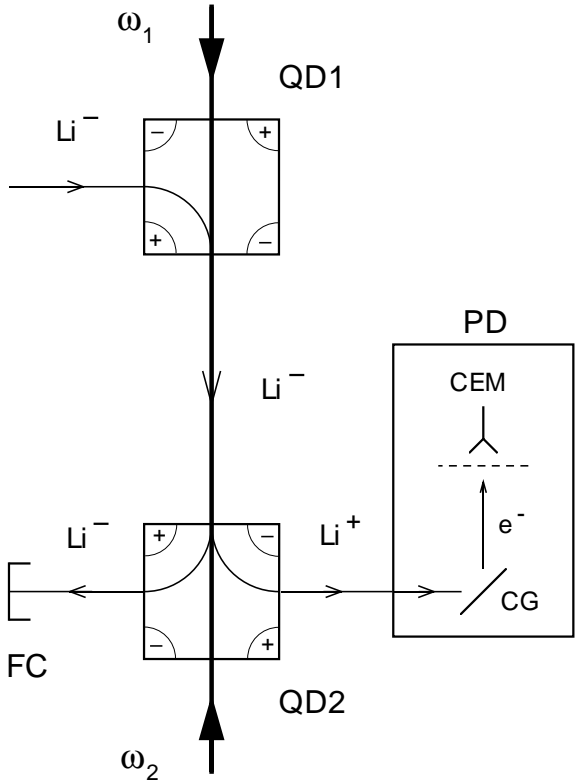, width=0.46\textwidth}
\caption{Interaction/detection chamber: {\sf QD1,QD2} {Electrostatic
quadrupole
deflectors}, {\sf CEM} {Channel electron multiplier},
{\sf PD} {Positive ion detector},
{\sf FC} {Faraday cup},
{\sf CG} {Conducting glass plate}. Ion- and laser beams were merged
in the interaction region
between the quadrupole deflectors over a common path of 0.5~m.}
\label{exp}
\end{figure}

Light of frequency $\omega_{1}$ was generated by 
a dye laser operated with Coumarine 307 
and light of frequency $\omega_{2}$ was produced by a dye laser 
operated with BMQ. Both dye lasers were pumped by a common 
XeCl excimer laser.
The pulse duration of both lasers was
about 15~ns. The maximum energy in a laser pulse delivered into the
interaction region  was 1.5~mJ for the radiation of frequency
$\omega_{1}$
and 200~$\mu$J for the radiation of frequency $\omega_{2}$.
During the experiment both lasers were attenuated, as will be discussed
below.

The two laser pulses counter propagated in the
interaction region and arrived simultaneously in the interaction
region. A shift of a few ns between
the pulses, occurring at both ends of the
interaction region, is negligible since the Li($2p\,^{2}\!P$) radiative
lifetime is 26.99(16)~ns \cite{MAl_95}.

The frequency $\omega_{1}$ was determined 
by combining Fabry-P\'erot fringes 
with optogalvanic spectroscopy. The Fabry-P\'erot fringes served
as frequency markers whereas transitions of Ne or Ar in a hollow cathode
lamp provided an absolute calibration of the energy scale.

\section{Results and discussion}
\label{results}

\subsection{Resonance ionization spectroscopy}
\label{ris}

\begin{figure}
\epsfig{file=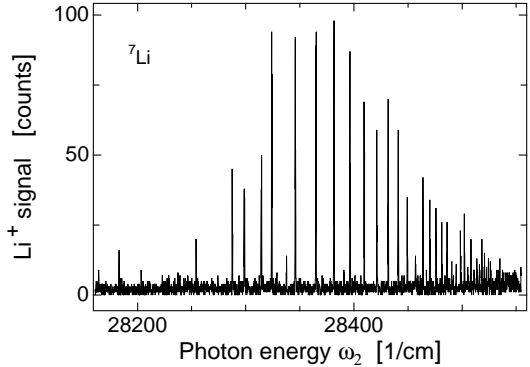, width=0.47\textwidth}
\caption{Rydberg series: Survey of the Rydberg states below the
ionization limit of the Li atom.
Note how the series are terminated at
28\,270~cm$^{-1}$. The lines below that limit we interpret as being
due to photoionization.
The energy scale is Doppler-shifted by -31~cm$^{-1}$.}
\label{uv2}
\end{figure}

First we had to find an appropriate resonance transition for the laser 
of frequency $\omega_{2}$.
A scan of $\omega_{2}$, during which $\omega_{1}$ is set to a frequency
far above the Li$(2p)$ threshold, is shown in Fig.\ \protect\ref{uv2}.
Below the ionization limit we found a wealth of lines. 
Due to strong saturation, the intensities of the lines are not
proportional to the corresponding transition probabilities; close to the ionization
limit the detection system became overloaded such that it only recorded
one count per laser shot thus attenuating the strongest lines. 
For most of the lines we also expect the
transition itself to be saturated. 
The Rydberg series
in Fig.\ \ref{uv2} exhibits a
cut off at 28\,270~cm$^{-1}$
indicating that the Rydberg atoms were mainly field ionized in an
electric field of about 200~kV/m in the second quadrupole deflector. 
We interpret the lines below the cut off as being due to
photoionization.

\begin{figure}
\epsfig{file=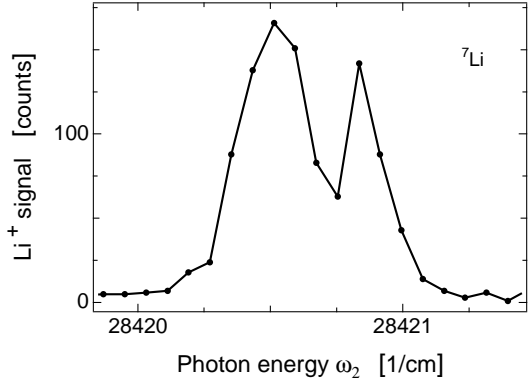, width=0.47\textwidth}
\caption{Fine structure: This spectrum of the
fine structure of the $2p\,^{2}\!P$ term is a magnification of the
spectrum depicted in Fig.\ {\protect\ref{uv2}}. The left peak
is due to the
transition from the $J=3/2$
level and the right one is due to the transition from
the $J=1/2$ level.
For the threshold
measurements, $\omega_{2}$ was tuned to the peak of the $J=1/2$
component.The energy scale is Doppler-shifted by -31~cm$^{-1}$ .}
\label{fein}
\end{figure}

A detailed scan of the transition used for the 
photodetachment threshold measurements is shown in Fig.\ \protect\ref{fein}.
We identified the two lines as
corresponding to transitions from the two fine structure
levels $2p\,^{2}\!P_{1/2,3/2}$ to the same Rydberg
state. The measured separation of 0.33(4)~cm$^{-1}$ between the two lines 
in Fig.\ \protect\ref{fein}
corresponds very well to the fine structure splitting of 0.337~cm$^{-1}$
of the $2p\,^{2}\!P$ term \cite{Moo_71_1}. 

The resolved fine structure shown in
Fig.\ \protect\ref{fein} demonstrates that the energy resolution of the
present
collinear beam apparatus is approximately 0.2~cm$^{-1}$, which is of 
the same order as the laser bandwidth.

\begin{figure}
\epsfig{file=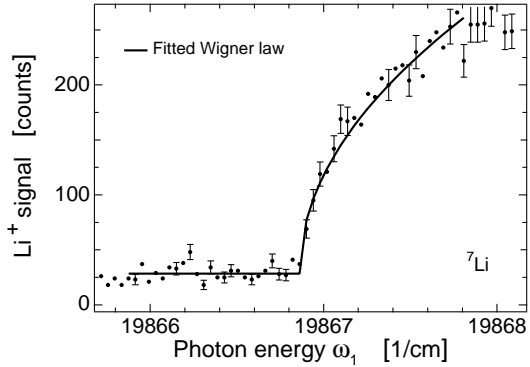, width=0.47\textwidth}
\caption{Li$(2p)+\epsilon s$ threshold: Measurement of the partial
relative
photodetachment cross-section of Li$^{-}$ around the Li($2p$)
threshold with anti-parallel laser (of frequency $\omega_{1}$)
and ion beam.
The solid line is a fit of the Wigner law to
the data in the range of the line. The error bars on selected data
points represent
the shot noise. Notice the very steep onset of the
photodetachment and the low background below the threshold. Each data
point is obtained from 100 laser pulses.}
\label{wig}
\end{figure}

\subsection{Electron affinity measurements}
\label{eameasurements}

A typical measurement of the Li(2$p$) photodetachment threshold with
anti-parallel laser and ion beams is shown in Fig.\ \ref{wig}. The laser 
frequency
$\omega_{2}$ was held constant and the intensity was set  to 
saturate the transition to the
Rydberg state.  The frequency $\omega_{1}$ was then
scanned over the Li($2p$) threshold.

It was established that two processes contributed to the background, 
namely (a): 
\begin{eqnarray}\label{stepone2}
\mbox{Li}^{-}(2s^{2}\,^{1}\!S) + \hbar\omega_{2} &\rightarrow &
\mbox{Li}(2p\,^{2}\!P) +
\mbox{e}^{-}\;,\nonumber\\
\mbox{Li}(2p\,^{2}\!P) + \hbar\omega_{2} &\rightarrow &
\mbox{Li}(ns,nd)\;,\nonumber\\ 
\mbox{Li}(ns,nd) &\leadsto & 
\mbox{Li}^{+}(1s^{2}\,^{1}\!S) + \mbox{e}^{-}\;,\nonumber
\end{eqnarray}
and (b) two-electron collisional ionization. 
Process (a)
produced nearly 80\% of the background, even though 
the intensity of the laser light with frequency $\omega_{2}$ was attenuated.
Process (b), two-electron
collisional ionization, contributed the remaining background at the
operating  pressure of $5\times 10^{-9}$~mbar ($5\times 10^{-7}$~Pa). 
This contribution was found to be
proportional to the pressure and is thus dominated by single collision 
double detachment, as has been discussed previously by Bae and Peterson
\cite{Bae_88}.

The laser intensities were too low for other processes, such as 
direct two-electron multiphoton detachment, to influence the
experiment. 
 
Close to the photodetachment threshold the cross section $\sigma (E)$ 
is well represented by the
Wigner law \cite{Wig_48} (Eq.~\ref{eqwig}).
If $E_{0}$ is the threshold energy and $E$ the photon energy, 
the Wigner law for electrons detached with
different angular momenta $l$ can be written as
 \begin{equation}
\label{eqwig}
\sigma(E) \propto \bigg(\sqrt{E-E_{0}}\,\bigg)^{2l+1}\;,\quad
\mbox{for}\;E\ge E_{0}\;.
\end{equation}
For the desired first step process in Eq.\ \ref{stepone},
both $s$- and $d$-wave final states for the detached electron are
allowed by parity and total angular momentum conservation.
At the threshold, however, the $s$-wave cross
section, starting with an infinite slope, completely dominates 
over the $d$-wave cross section.

We fitted the Wigner law (Eq.\ \ref{eqwig}) for $s$-wave detachment ($l=0$)
to the recorded positive ion signal. The data shown in Fig.\ \ref{wig} was
recorded with
the laser ($\omega_{1}$) anti-parallel to the ion beam.
 The fit window
is the range of the plotted line. The fit yields
 a value
for the red-shifted threshold energy, $E_{0}^{r}$ .
In order to eliminate the Doppler shift we repeated the measurements using
parallel laser and ion beams to determine the  blue-shifted threshold
energy,
 $E_{0}^{b}$ 
The threshold energy corrected for the Doppler to all orders, $E_{0}$, is given
by the
geometric mean of the two measurements: 
\begin{eqnarray}
\label{schwelle}
E_{0} &=& \sqrt{E_{0}^{b} E_{0}^{r}}\\
&=& 19\,888.55(16)\;\mbox{cm$^{-1}$}\quad .\nonumber
\end{eqnarray}

The interaction of the induced dipole moment of the atom with the
electron will limit the range of validity of the Wigner law.
O'Malley \cite{OMa_65} obtained an expression 
which accounts for the residual induced-dipole point-charge interaction,
\begin{equation}
\label{omalley}
\sigma(k) \propto k^{2l+1}
\left[1-\frac{4\alpha k^{2}\ln k}{(2l+3)(2l+1)(2l-1)} + 
{\cal{O}}(k^2)\right]\quad ,
\end{equation}
where $k=\sqrt{2(E-E_{0})}$ and $\alpha $ is in atomic units.
This equation can be used to estimate an
approximate range of validity for the Wigner law by determining the
conditions under which the second term, and therefore higher terms, become
negligible. With a dipole
polarizibility of $\alpha = 152~a.u.$ for the Li $2p\,^{2}\!P$ state 
\cite{Man_75}, the second term is much less than unity  up to 1~cm$^{-1}$
 above the 
threshold, and 
consequently the Wigner law should  be applicable over this range.
This, of course, assumes that there are no resonances in this region, a
condition
which 
has been previously established by Dellwo et al.\cite{Del_92_2}.

Saturation of the  signal can originate in the
photodetachment process or the detection system. In the data shown in
Fig.\ \ref{wig} the detection system  starts to
saturate at about 250 counts (for 100 laser pulses). We attenuated the
intensity of the laser of frequency $\omega_{1}$ sufficiently to avoid any
saturation of the signal within the first 1.0~cm$^{-1}$ above the
threshold.

To determine the electron affinity of Li we had to subtract the well known
$2p\,^{2}\!P_{1/2} \rightarrow 2s\,^{2}\!S$ transition energy 
of 14\,903.648\,130(14)~cm$^{-1}$ \cite{Moo_71_1} from the measured
Doppler-
\begin{table}
\protect\caption{Comparison of different measurements and
calculations of the Li
electron affinity.}
\label{tab1}
\begin{tabular}{lc}
Author & Affinity in meV \\
\tableline
{\bf Experiment}:&\\
This work & $618.049\pm 0.020$\\
Feldmann (1976)\cite{Fel_76} & $618.2\pm 0.5$\\
Bae and Peterson (1985)\cite{Bae_85} & $617.3\pm 0.7$\\
Dellwo et al.~(1992)\cite{Del_92_2} & $617.6\pm 0.2$\\
{\bf Theory} (after 1992):&\\
Chung and Fullbright (1992)\cite{Chu_92_2} & $617.4\pm 0.2$\\
Froese Fischer (1993)\cite{Fis_93} & 617.64
\end{tabular}
\end{table}
\begin{figure}
\epsfig{file=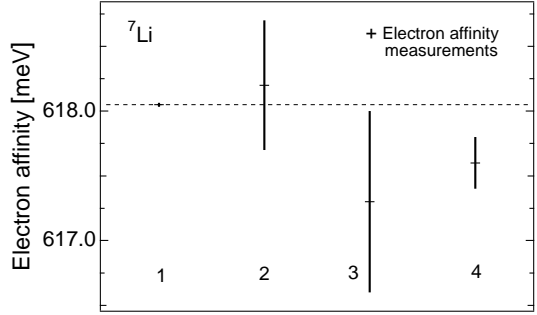, width=0.47\textwidth}
\caption{Electron affinity: Graphical comparison
of the different
experimental lithium electron affinity values with their respective
uncertainties. From left to right the
values are: {\bf\sf 1}~This work, 
{\bf\sf 2}~Feldmann,
{\bf\sf 3}~Bae et.~al and
{\bf\sf 4}~Dellwo et.~al.~. For references see Tab.\
\protect\ref{tab1}.}
\label{EAlithium}
\end{figure}
corrected threshold energy $E_{0}$ (Eq.\ \protect\ref{schwelle}).

We determined the Li electron
affinity to be 4\,984.90(17)~cm$^{-1}$. To the error there are two major
contributions: 0.13~cm$^{-1}$ is due to a calibration uncertainty of
the hollow cathode lamps, and the rest is due to statistical 
scattering of the fitted threshold values. 

In order to convert our electron affinity in cm$^{-1}$ to eV we used the
recommended factor of (1/8\,065.5410)
$\big[\mbox{eV}/(\mbox{cm$^{-1}$})\big]$ \protect\cite{Coh_88} and obtained
0.618\,049(20)~eV.
In Fig.\ \protect\ref{EAlithium} this value is compared with previous 
experiments.
These values are also compiled in Tab.\ \protect\ref{tab1} together with
some 
recently calculated Li electron affinities. 

\section{Summary and conclusion}
\label{summary}

We have demonstrated that photodetachment spectroscopy combined with
resonance ionization is a powerful method for studying partial
photodetachment cross sections of negative ions.
The collinear beam geometry simultaneously provides high
sensitivity, due to the large interaction volume, and excellent
resolution, due to velocity compression.  I addition, removal of the Doppler
shift to all orders can be achieved  by use of two separate measurements involving 
parallel and anti-parallel laser and ion beams. Both 
merits can be fully
retained for channel specific photodetachment investigations with the
excitation scheme used in this experiment. Our improved Li electron
affinity of 0.618\,049(20)~eV reveals the potential of this method, and this
method can, in principle, 
 be extended to  essentially all  elements.

We are currently developing this type of
state-selective detection scheme in connection with on-going
studies of high lying doubly excited states of the Li$^{-}$ ion
\cite{Ber_95_1}. 

\section{Acknowledgments}

Financial support for this research
has been obtained from the Swedish Natural Science
Research Council (NFR). 
Personal support was received from the Wenner-Gren Center Foundation for
Igor Kiyan. David Pegg acknowledge the support from the Swedish Institute
and  and the US
Department of Energy, Office of Basic Energy Sciences, Division of Chemical Sciences.

\end{document}